\documentclass[pdflatex,sn-nature]{sn-jnl}

\usepackage{lineno}
\usepackage{xr}
\usepackage{graphicx}%
\usepackage{multirow}%
\usepackage{amsmath,amssymb,amsfonts}%
\usepackage{amsthm}%
\usepackage{mathrsfs}%
\usepackage[title]{appendix}%
\usepackage{xcolor}%
\usepackage{textcomp}%
\usepackage{manyfoot}%
\usepackage{booktabs}%
\usepackage{algorithm}%
\usepackage{algorithmicx}%
\usepackage{algpseudocode}%
\usepackage{listings}%
\usepackage{subcaption}
\usepackage[inkscapelatex=false]{svg}

\usepackage{soul} 

\theoremstyle{thmstyleone}%
\theoremstyle{thmstyletwo}%

\theoremstyle{thmstylethree}%

\begin{document}
\title[Article Title]{Diffusion-based Probabilistic Air Quality Forecasting with Mechanistic Insight}


\author[1,2,3,4]{\fnm{Ao} \sur{Ding}}\email{ao2024.ding@connect.polyu.hk}
\author*[3,4]{\fnm{Aoxing} \sur{Zhang}}\email{zhangax@sustech.edu.cn}
\author*[3,4,5]{\fnm{Tzung-May} \sur{Fu}}\email{fuzm@sustech.edu.cn}
\author*[1]{\fnm{Yuanlong} \sur{Huang}}\email{yhuang@eitech.edu.cn}
\author*[2]{\fnm{Qianjie} \sur{Chen}}\email{qianjie.chen@polyu.edu.hk}
\author[3,4]{\fnm{Yuyang} \sur{Chen}}
\author[3,4]{\fnm{Jiajia} \sur{Mo}}
\author[3,4]{\fnm{Wei} \sur{Tao}}
\author[3,4]{\fnm{Wai-Chi} \sur{Cheng}}
\author[3,4]{\fnm{Lei} \sur{Zhu}}
\author[3,4]{\fnm{Xin} \sur{Yang}}
\author[6]{\fnm{Guy} \sur{Brasseur}}

\affil[1]{School of the Environment and Sustainable Engineering, Eastern Institute of Technology, Ningbo, Zhejiang, 315200, China}
\affil[2]{Department of Civil and Environmental Engineering, The Hong Kong Polytechnic University, Hong Kong SAR, China}
\affil[3]{State Key Laboratory of Soil Pollution Control and Safety, Shenzhen Key Laboratory of Precision Measurement and Early Warning Technology for Urban Environmental Health Risks, School of Environmental Science and Engineering, Southern University of Science and Technology, Shenzhen, Guangdong, 518055, China}
\affil[4]{Guangdong Provincial Field Observation and Research Station for Coastal Atmosphere and Climate of the Greater Bay Area, Southern University of Science and Technology, Shenzhen, Guangdong, 518055, China}
\affil[5]{National Center for Applied Mathematics Shenzhen, Shenzhen, Guangdong, 518055, China}
\affil[6]{Max Planck Institute for Meteorology, Hamburg, 20146, Germany}

\abstract{
Current operational air quality forecasts are computationally expensive, sensitive to errors in physics and emissions, and often neglect weather-related uncertainty. To address these limitations, we present AirFusion, a hybrid, diffusion-based framework that synergistically integrates knowledge from chemical transport models with real-world observational constraints to enable accurate and efficient probabilistic regional air quality prediction. We apply AirFusion to generate operational 6-day, 30-member ensemble forecasts of surface ozone across China, initialized with observations and driven by ensemble weather forecasts. AirFusion outperforms existing operational benchmarks, achieving substantially lower forecast errors against surface measurements, while also providing ensemble-based diagnostics that explicitly quantify the impacts of weather uncertainty on air quality predictability. Moreover, AirFusion can rapidly adapt to evolving emissions through fine-tuning with only one month of recent observations. These attributes establish AirFusion as a powerful and extensible framework for next-generation probabilistic air quality forecasting, with clear potential for application to other pollutants and regions.
}




\maketitle

\section{Introduction}\label{sec1}

Accurate and timely air quality forecasting is essential for effective air pollution management, guiding critical actions such as implementing emission controls and issuing public health advisories. Current operational air quality forecasts predominantly rely on physics-based chemical transport models (CTMs), which predict pollutant concentrations by solving complex sets of differential equations for atmospheric physics and chemistry, using online or offline meteorological forecasts and prescribed pollutant emission inventories as inputs.  However, CTM-based forecasts are highly susceptible to uncertainties from several sources: inaccuracies in emission inventories  \cite{napelenok2011dynamic}, imperfect representations of multi-scale atmospheric processes \cite{russell2000narsto}, and innate uncertainties in the input meteorological forecasts \cite{zhang2007impacts}. While the meteorological uncertainty could, in principle, be quantified by driving CTMs with multiple members of an ensemble weather forecast \cite{zhang2023deep}, that approach is computationally prohibitive and generally not employed in daily operations.

Recent advances in artificial intelligence (AI), particularly the demonstrated success of AI-based global weather forecasting \cite{lam2023learning, price2025probabilistic, li2024generative, bodnar2025foundation}, provide a transferable foundation for AI-based air quality (AI-AQ) forecasting. Several studies have developed observation-driven AI-AQ forecast models, progressing from forecasting individual locations to regional and global domains \cite{bodnar2025foundation, siwei2025enhancing}. These observation-driven models learn the meteorology-pollution relationships directly from historical observations, bypassing the need for explicit emission inventories and mechanistic representation. A key advantage of the AI-AQ paradigm is its computational efficiency, which enables rapid ensemble predictions to account for weather forecast uncertainty \cite{zhang2023deep}.

However, unlike AI-based weather forecasting models, which are trained on extensive, physically consistent meteorological reanalyses, observation-driven AI-AQ forecasting is severely limited by the scarcity and non-stationarity of observations. For example, although hourly pollutant measurements from over 1000 surface sites across China have been available since 2013, the underlying meteorology-pollution relationships within these data are non-stationary, as emissions have been continually reshaped by policies and technologies \cite{wen2024combined}. Long-term monitoring data in the United States and Europe have been similarly affected by evolving sectoral activities, control technologies, and regulations\cite{mcduffie2020global}. In many other regions, systematic surface monitoring data are unavailable. As a result, current observation-driven AI-AQ models exhibit relatively low and unstable forecast accuracy. For instance, even when driven by meteorological reanalysis, Microsoft's Aurora model produced 00 UTC surface ozone hindcasts with a normalized mean bias (NMB) of 28\% over China \cite{bodnar2025foundation}. This performance shortfall highlights the critical need for methodological advances that can compensate for sparse and non-stationary training data in AI-AQ forecasting.

We propose a hybrid, diffusion-based framework to address the dual challenges of inaccuracies in CTMs and the scarcity and non-stationarity of observational data. This framework operates in two stages: pre-training on comprehensive CTM simulations to learn physically robust meteorology-air quality relationships, followed by fine-tuning on a small set of recent observations to correct biases, adapt to latest emission changes, and align with real-world conditions. Diffusion models are well-suited to this hybrid paradigm, as they learn data generation by progressively reversing a noise-adding process \cite{ho2020denoising}, which naturally separates feature-learning by scale: early denoising steps recover broad patterns associated with synoptic weather and regional pollution, while later steps refine mesoscale and local details. This multi-scale learning aligns with the hierarchy of processes that shape air quality. Beyond scale-awareness, diffusion models also offer algorithmic advantages over other deep learning architectures for environmental prediction. Diffusion models dynamically capture evolving spatial dependencies, unlike Graph Neural Networks (GNNs) and Convolutional Neural Networks (CNNs) that rely on static adjacency matrices \cite{harshvardhan2020comprehensive}. Also, diffusion frameworks exhibit greater training stability than Generative Adversarial Networks (GANs), avoiding mode collapse \cite{goodfellow2014generative}. The efficacy of diffusion models for environmental modeling is evidenced by their recent successful application in high-resolution weather forecasting \cite{price2025probabilistic} and cloud reanalysis\cite{xiao2025high}.

We present AirFusion, a diffusion-based AI-AQ forecasting framework that synergistically learns from CTM simulations and observational data. Applied to produce operational 6‑day, 3‑hourly ensemble surface ozone forecasts over China, AirFusion demonstrates superior skill compared to existing operational systems. We examine how the model learns multi‑scale features from both CTM outputs and real‑world observations, evaluate its rapid adaptability to changing pollutant emissions, and explicitly quantify the critical impact of meteorological uncertainty on forecast accuracy. Through ensemble‑based diagnostics, AirFusion provides reliable probabilistic forecasts of ozone exceedance across Chinese cities. This work demonstrates AirFusion as a significant advance over traditional CTMs and prior AI‑AQ models in terms of forecast accuracy, spatial resolution, computational efficiency, and operational robustness, offering a powerful and extensible tool for managing air‑pollution impacts on public health and ecosystems.



\section{Overview of the AirFusion ensemble air quality forecast system}

\begin{figure}
    \centering
    \includegraphics[width=\linewidth]{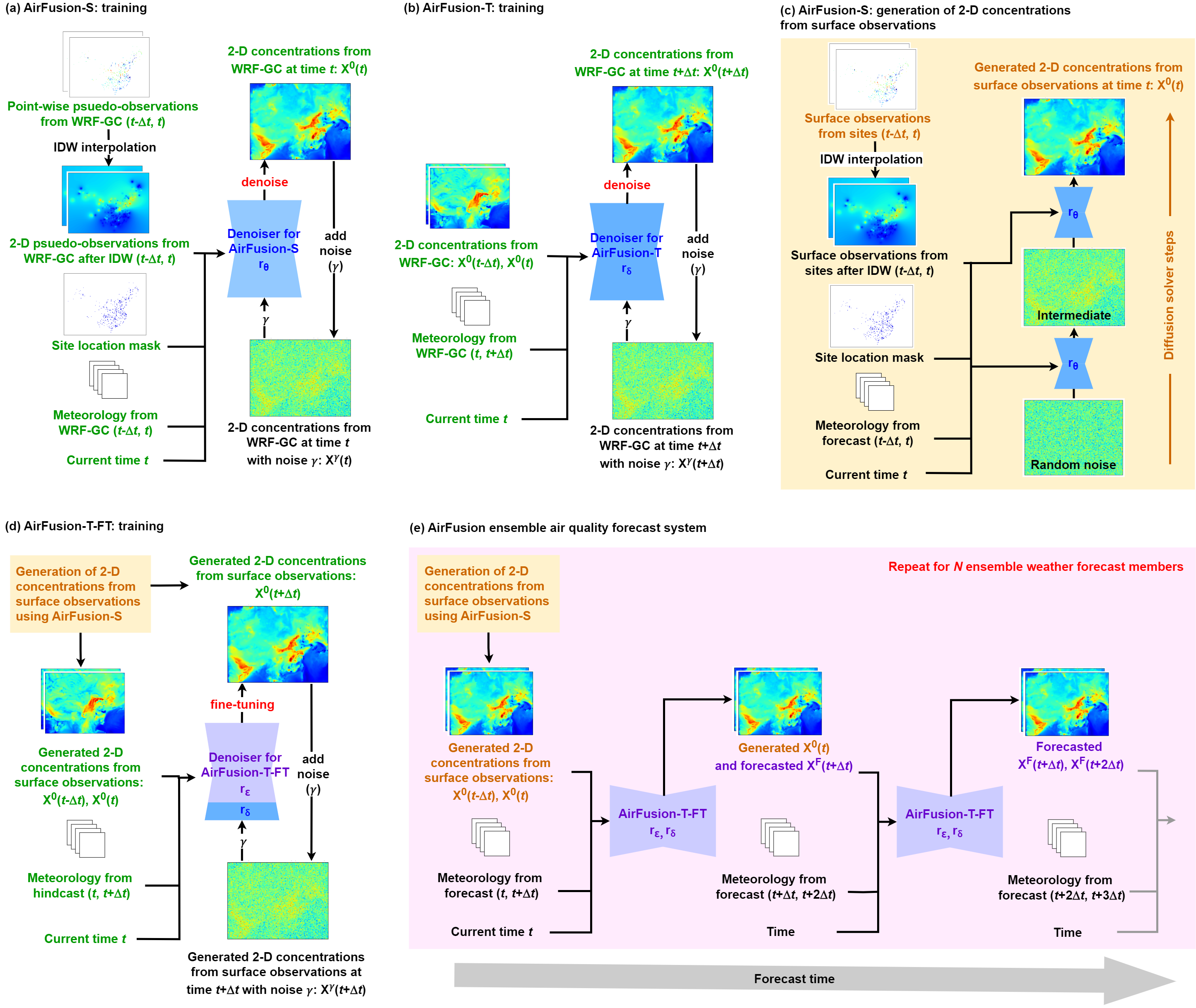}
    \caption{
\textbf{Schematic of the AirFusion ensemble air quality forecast system.} \textbf{a},\textbf{b},\textbf{e}, Training of AirFusion's three modules: AirFusion-S (\textbf{a}), AirFusion-T (\textbf{b}), and AirFusion-T-FT (\textbf{e}). \textbf{c}, Given surface observations of pollutant concentrations, AirFusion-S generates a continuous 2-D pollutant concentration field, $X^0(t)$. \textbf{e}, In operational forecast mode, AirFusion-S generates initial conditions, $X^0(t_0-\Delta t)$ and $X^0(t_0)$, from surface pollutant observations. Weather forecasts are then input to AirFusion-T-FT to predict surface pollutant concentrations at the next time step, $X^F(t_0+\Delta t)$. Subsequent forecasts use the AirFusion-T-FT outputs and ensemble weather forecasts to advance in time through the desired forecast period. The above procedure is then repeated for all $N$ members of an ensemble weather forecast. }
    \label{fig_schematic}
\end{figure}

Fig.  \ref{fig_schematic} shows the schematic of the AirFusion ensemble air quality forecast system, which consists of three specialized modules: AirFusion-S, AirFusion-T, and AirFusion-T-FT. All three modules are built upon a common diffusion model backbone (\textit{Methods}). The AirFusion-S module interpolates spatially sparse observations from the previous and present time steps to generate a continuous 2-D surface pollutant concentration field for the present time, $X^0(t)$ (Fig.  \ref{fig_schematic}a). AirFusion-T advances 2-D surface pollution concentration fields from the previous and present time steps to the next time step, driven by meteorological fields at the present and next time steps (Fig.  \ref{fig_schematic}b). Both AirFusion-S and AirFusion-T acquire latent spatiotemporal meteorology-pollution relationships through training on outputs from a physics-based CTM, in our case the WRF-GC model (\textit{Methods}). Finally, the AirFusion-T-FT module is created by fine-tuning AirFusion-T with recent observations and meteorological forecasts; this process corrects systematic biases in the CTM simulations, calibrates the meteorological differences between the CTM and the weather forecast model, and adjusts to current emission levels, enabling AirFusion-T-FT to more accurately propagate pollutant concentrations with time (Fig.  \ref{fig_schematic}d, \textit{Methods}).

In forecast mode (Fig.  \ref{fig_schematic}e), AirFusion-T-FT generates a continuous 2-D surface pollutant concentration image for the first forecast time step ($X^F(t_0+\Delta t)$), driven by initial conditions of pollutant concentrations generated by AirFusion-S from observations at times $t_0-\Delta t$ and $t_0$ (Fig.  1c) and weather forecasts for times $t_0$ and $t_0+\Delta t$. Subsequent forecasts use the model's outputs to step forward in time through the desired forecast period. The entire time-stepping procedure is then repeated for each of the $N$ members of an ensemble weather forecast. In the end, AirFusion produces an $N$-member ensemble forecast of 2-D pollutant concentration sequences; each member represents air quality evolution under a possible future weather realization. For any given location, the mean of the $N$ ensemble members represents the most probable pollutant concentration prediction, while the ensemble distribution manifest the range of possible air quality outcomes due to weather forecast uncertainty.

\section{Application of AirFusion to forecast surface ozone over China}

\begin{figure*}[htp]
    \centering
        \includegraphics[width=\linewidth]{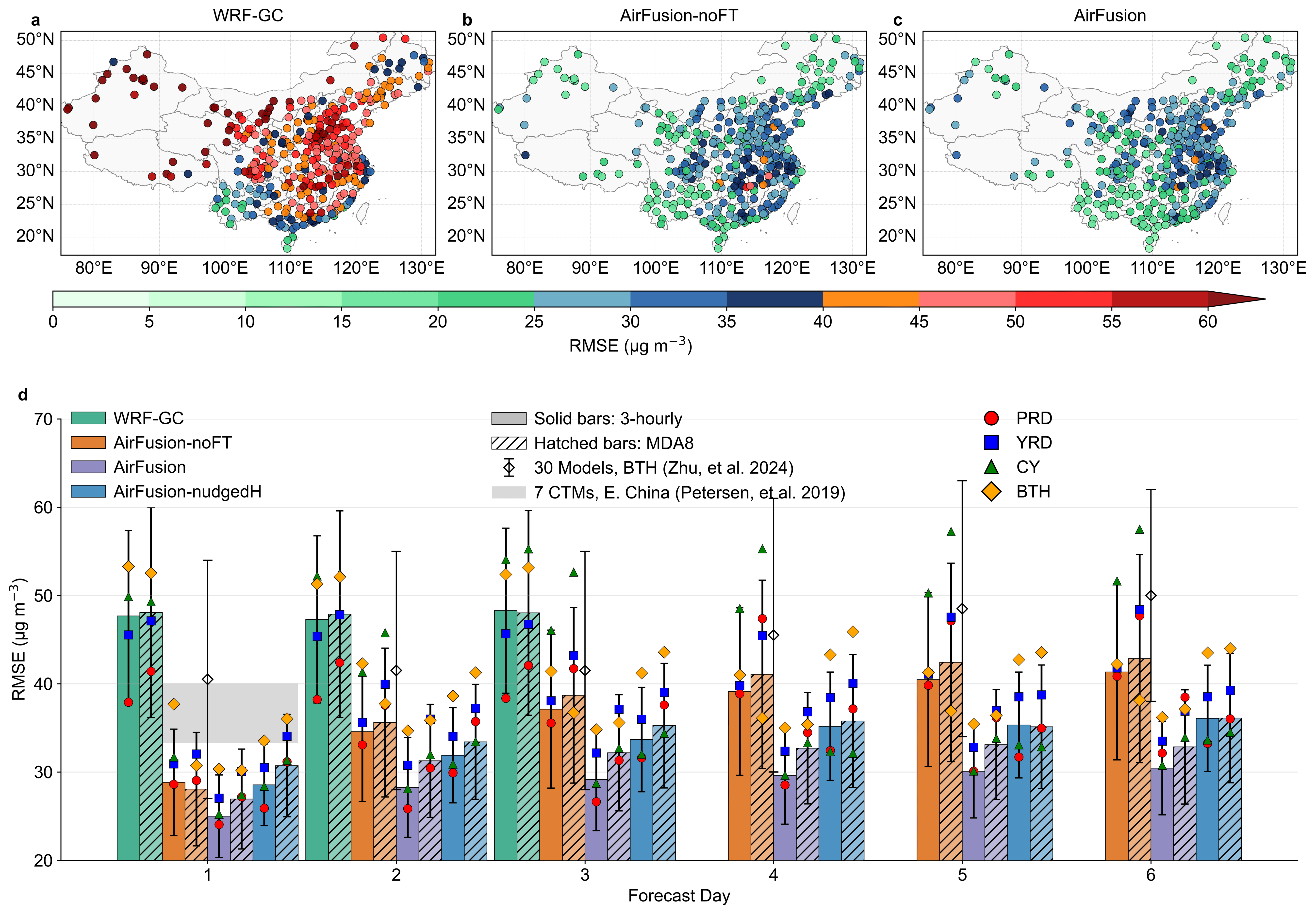}    
    \caption{
\textbf{Performance of the AirFusion system in forecasting surface ozone concentrations across China.} \textbf{a-c}, Root mean square errors (RMSEs) of Day-1 MDA8O\textsubscript{3} forecasts for July to September 2024 from three models: WRF-GC (\textbf{a}), AirFusion without fine-tuning (AirFusion-noFT) (\textbf{b}), and the fully fine-tuned AirFusion system (\textbf{c}). MDA8O\textsubscript{3} forecasts in AirFusion-noFT and AirFusion were derived by linearly interpolating their respective 3-hourly ensemble mean ozone forecasts. \textbf{d}, Comparisons of RMSEs of MDA8O\textsubscript{3} (solid bars) and 3-hourly ozone (hatched bars) concentrations averaged across 341 Chinese cities for Day 1 to Day 6 lead times. Green, WRF-GC; orange, AirFusion-noFT; purple, AirFusion; blue, AirFusion-nudgedH hindcast. Black whiskers indicate standard deviations across Chinese cities. Also shown are the RMSEs over four major megacity clusters (colored symbols): Beijing–Tianjin–Hebei (BTH), Yangtze River Delta (YRD), Pearl River Delta (PRD), and Sichuan–Chongqing (CY), along with the range of RMSEs from previous multi-model assessments over the BTH \cite{zhu2024comparative} (black diamond and whiskers) and ensemble model over eastern China\cite{petersen2019ensemble} (grey shaded area).}

    \label{fig:map_rmse}
\end{figure*}

We applied the AirFusion framework to generate 6-day surface ozone concentration forecasts across China (domain shown in Fig.  \ref{fig:map_rmse}, $72.8^\circ$--$137.2^\circ$E, $16.4^\circ$--$54.7^\circ$N) at 27-km spatial resolution for July to September 2024. The AirFusion-S and AirFusion-T modules were trained on WRF-GC simulations for June to September 2021 (\textit{Methods}). To best reflect the emission levels and real-world conditions relevant to the forecast period, the AirFusion-T-FT module was fine-tuned using surface ozone measurements across China (\url{https://www.cnemc.cn}, last accessed: 20 Jan 2026) and weather forecasts from the preceding month (June 2024; \textit{Methods}). In this forecast application, we drove AirFusion with the 30-member NOAA GEFS ensemble weather forecast with 3-hourly temporal resolution (\textit{Methods}), which also set the output temporal resolution of AirFusion. The system was initialized daily shortly after 23 China Standard Time (CST) (15 UTC) of Day 0, with the first valid forecast issued for 02 CST the following day (i.e., Day 1). AirFusion-S generated initial conditions of pollutant concentrations by spatially interpolating surface ozone observations from 20 and 23 CST of Day 0.  All necessary inputs, including surface ozone observations and ensemble weather forecasts, were available operationally from their respective sources at the time of initialization.

Fig.  \ref{fig:map_rmse}, S1, and S2 compare the performance of the fully fine-tuned AirFusion system against the WRF-GC model, other CTM-based and statistical models, and an AirFusion implementation without fine-tuning (hereafter AirFusion-noFT, consisting of AirFusion-S and AirFusion-T) for surface ozone forecasts across 341 Chinese cities at different lead times. The root-mean-square error (RMSE) of WRF-GC's maximum daily 8-h average O\textsubscript{3} (MDA8O\textsubscript{3}) concentration forecasts exceeded 40 $\mu$g m\textsuperscript{-3} for most cities, particularly over high-ozone regions such as the Beijing–Tianjin–Hebei (BTH) area and western China (Fig.  \ref{fig:map_rmse}a). The national mean RMSE of WRF-GC was 48.1$\pm$11.9 $\mu$g m\textsuperscript{-3} on Day 1 and remained elevated through Day 3, comparable to the RMSEs of previous CTM-based or statistical summertime ozone forecasts over the BTH area (27 to 59 $\mu$g m\textsuperscript{-3} on Day 1, increasing to 37 to 63 $\mu$g m\textsuperscript{-3} on Day 6) \cite{zhu2024comparative} and eastern China (38 to 80 $\mu$g m\textsuperscript{-3} for 72-h forecasts) \cite{petersen2019ensemble}. WRF-GC also showed widely scattered normalized mean bias (NMB) values of 4.9\%$\pm$29.2\% and low temporal correlations ($r$ = 0.47$\pm$0.29) against surface observations across Chinese cities (Fig.  S1 and S2). These diagnostics showed that WRF-GC and other current CTM-based forecast systems are regionally biased and limited in their ability to capture the spatiotemporal evolution of observed surface ozone across Chinese cities. 

In comparison, AirFusion-noFT showed substantially improved performance in forecasting MDA8O\textsubscript{3} on Day 1, with a smaller national mean RMSE of 28.1$\pm$6.4 $\mu$g m\textsuperscript{-3}, less scattered NMBs (6.2\%$\pm$15.5\%), and better temporal correlations ($r$ = 0.69$\pm$0.11). These improvements demonstrated that initializing ozone forecasts with observed concentrations via AirFusion-S partially mitigated the systematic biases inherited from WRF-GC simulations. However, without fine-tuning, the national mean forecast RMSEs increased by approximately 50\% and correlations declined to 0.44$\pm$0.14 from Day 1 to Day 6, reflecting the fading influences of observationally constrained initial conditions and the accumulation of errors from both the CTM and weather forecasts at longer lead times. 

The fine-tuned AirFusion system vastly outperformed WRF-GC, AirFusion-noFT, and other CTM-based or statistical surface ozone forecasts over China (Fig.  \ref{fig:map_rmse}, S1, and S2). On Day 1, AirFusion exhibited markedly lower RMSEs (26.9$\pm$5.7 $\mu$g m\textsuperscript{-3}), lower and less scattered NMBs (1.4\%$\pm$13.4\%), and higher temporal correlations ($r=$ 0.7$\pm$0.1) across Chinese cities and over four major Chinese megacity clusters. Furthermore, AirFusion's forecast skills showed minimal degradation with lead time: the national mean RMSE of its Day 6 MDA8O\textsubscript{3} forecast was 32.8 $\mu$g m\textsuperscript{-3}, only 22\% higher than the Day 1 value. In particular, AirFusion substantially reduced the ozone forecast errors over the Sichuan-Chongqing (CY) area in Southwest China that were evident in the other models. Air quality forecasting challenges in the CY area have been well-documented due to its complex terrain \cite{tao2014formation}, poor cloud representation \cite{yahya2016decadal}, and rapidly changing emissions \cite{zhao2018spatial}. AirFusion's enhanced performance stemmed from fine-tuning on only five-month's worth of surface ozone observations, effectively correcting errors in the temporal evolution of ozone pollution, while also preserving the mechanistic foundation provided by WRF-GC. 

We also compared the performance of AirFusion's surface ozone forecasts across China for July 2024 against the 5-day, 0.4\textdegree ozone hindcasts at 00 UTC from Microsoft's Aurora model \cite{bodnar2025foundation}. Aurora's hindcasts were driven by meteorological reanalysis and thus less affected by the inherent uncertainty of weather forecasts. Nevertheless, AirFusion produced more accurate surface ozone forecasts throughout the lead times, with lower RMSE and NMB values across Chinese cities (Fig. S3).

The superior forecast accuracy and computational efficiency of AirFusion position it as a powerful tool for air quality management in China. While AirFusion's ozone forecast RMSEs did not consistently meet the country's operational forecast accuracy targets ($\pm$12 $\mu$g m\textsuperscript{-3} for MDA8O\textsubscript{3} <160 $\mu$g m\textsuperscript{-3}, and $\pm$16.5 $\mu$g m\textsuperscript{-3} for MDA8O\textsubscript{3} $\ge$160 $\mu$g m\textsuperscript{-3}) for all cities, we found that its residual errors were predominantly driven by the uncertainties in the meteorological inputs. To quantify this source of error, we constructed an ozone hindcast system with minimal uncertainty (AirFusion-nudgedH) by fine-tuning with surface ozone observations and down-scaled meteorological reanalysis for June to September 2021 (\textit{Methods}). We then used AirFusion-nudgedH to hindcast surface ozone during July to September 2024, driven by down-scaled meterological reanalysis. Surprisingly, the AirFusion-nudgedH hindcasts showed poorer skills than the AirFusion forecasts (Figs. \ref{fig:map_rmse} ), indicating that the meteorological reanalysis itself failed to accurately capture the multi-scale meteorology relevant to surface ozone pollution \cite{manney2017reanalysis,li2023intercomparison}.  In operational AirFusion forecasts, this meteorological representation error was partially mitigated by the multiple ensemble members, such that the ensemble mean more probably captured the actual meteorological condition. This finding highlighted the necessity of accounting for weather uncertainty in operational air quality forecasting and also suggested that China's current targets for forecast accuracy potentially warrant revision, because these targets may not be attainable for all cities with existing weather forecast technologies. In terms of computational efficiency, AirFusion produced a one-day (27-km spatial resolution, 3-hourly temporal resolution) 30-member surface ozone ensemble forecast for all of China on a single RTX 4090 GPU in 40 seconds, more than four orders of magnitude faster than WRF-GC's one-day (27-km spatial resolution, hourly temporal resolution), single-member forecast simulated on 80 CPU cores.

\section{Multi-scale feature learning of AirFusion}
\begin{figure}
    \centering
    \includegraphics[width=\linewidth]{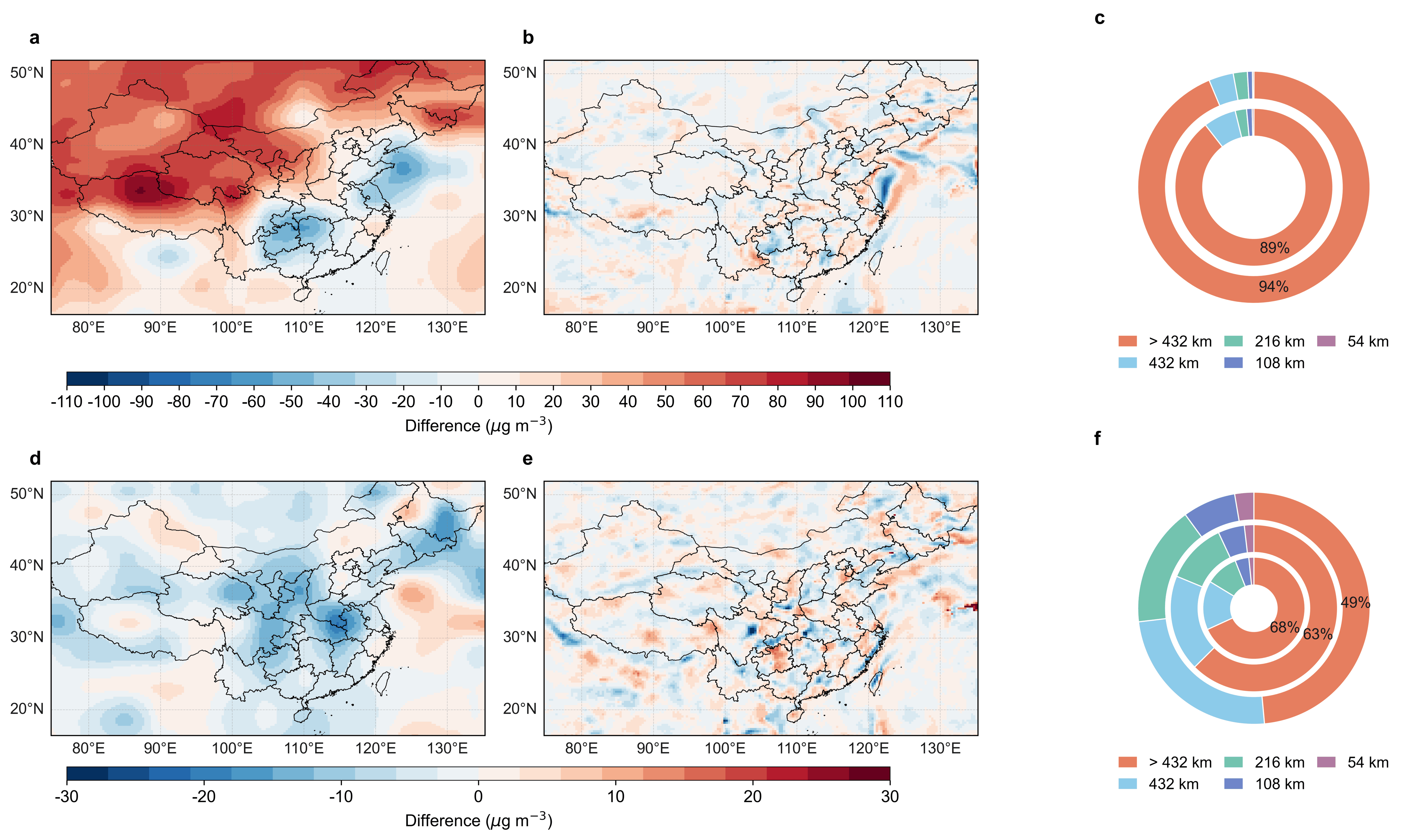}
    \caption{\textbf{Spatial scales of surface ozone pollution features learned by the AirFusion system.} \textbf{a-c}, Two-dimensional discrete wavelet decomposition of forecasted MDA8O\textsubscript{3} concentration differences between WRF-GC and AirFusion-noFT on Day 1 at scales $\ge$432 km (\textbf{a}) and <432 km (\textbf{b}), and the energy distributions on Days 1 (outer) and 3 (inner) (\textbf{c}). \textbf{d-f}, Two-dimensional discrete wavelet decomposition of forecasted MDA8O\textsubscript{3} concentration differences between AirFusion and AirFusion-noFT on Day 1 at scales $\ge$432 km (\textbf{d}) and <432 km (\textbf{f}), and the energy distributions on Days 1 (outer), 3 (mid), and 5 (inner) (\textbf{f}).}
    \label{fig:wavelet}
\end{figure}

To better understand how the AirFusion system learned the spatiotemporal features of surface ozone pollution to generate accurate forecasts, we performed a two-dimensional discrete wavelet decomposition (\textit{Supplementary Information}) on the forecast MDA8O\textsubscript{3} differences between AirFusion-noFT and WRF-GC (Fig.  \ref{fig:wavelet}a-c), and those between AirFusion and AirFusion-noFT (Fig.  \ref{fig:wavelet}d-f). Relative to WRF-GC, AirFusion-noFT predicted higher ozone concentrations over western China and lower concentrations over southern China. More than 89\% of the total energy in the forecast differences between the two models resides in spatial scales $\ge$432 km, with very little contribution from smaller scales. This dominance of difference energy by the large scale indicated that initializing with observed surface ozone concentrations effectively corrected WRF-GC's systematic biases over broad regions, leading to the smaller and less variable RMSEs and NMBs of AirFusion-noFT (Fig.  \ref{fig:map_rmse} and S2).

In contrast, the bias-correction in AirFusion relative to AirFusion-noFT occurred at finer spatial scales. The differences between the two models arose from fine-tuning AirFusion with recent surface ozone observations, which improved AirFusion-T-FT's ability to accurately advance ozone concentrations in time. The largest corrections were over central and eastern China (Fig. 3d-f) reflecting the stronger observational constraints available in these densely monitored regions. More than half (51\%) of the total difference energy resided at meso- and small scales on Day 1, a fraction that remained substantial (32\%) even at Day 5. This persistence indicated that fine-tuning enhances AirFusion's overall representation of key meso- small-scale processes throughout the forecast period, such as emissions, boundary-layer mixing, local photochemistry, and other meteorological influences. The resulting improvement in temporal representation explained why AirFusion's forecast skills degraded more slowly at longer lead times than AirFusion-noFT (Fig.  \ref{fig:map_rmse}, S1, and S2).

\section{Fast and flexible adaptation of AirFusion to changes in pollutant emissions}
\label{chap:fast_flex}
\begin{figure}
    \centering
    \includegraphics[width=0.7\linewidth]{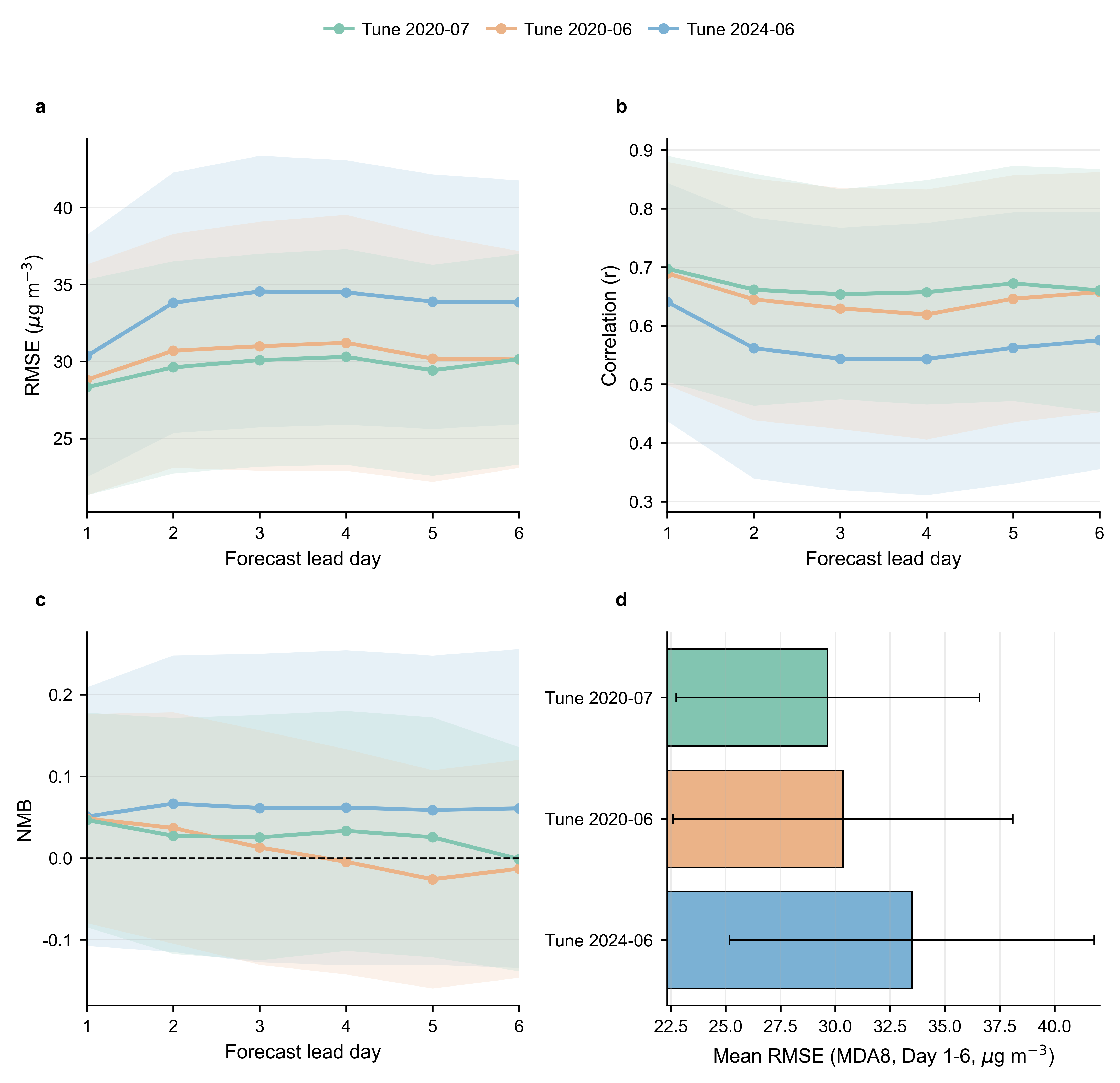}
    \caption{Comparison of AirFusion's performance in predicting surface ozone of August 2020 when fine-tuned with observations in June 2024 (blue), June 2020 (orange), and July 2020 (green). \textbf{a-c,} RMSEs (\textbf{a}), temporal correlations (\textbf{b}), and NMBs (\textbf{c}) for Day 1 to Day 6. Solid lines and shaded areas indicate the national means and standard deviations across 341 Chinese cities, respectively. \textbf{d}, mean RMSEs for Days 1 to 6.}
    \label{fig:flex_adaptation}
\end{figure}

A key advantage of AirFusion is its ability to rapidly adapt to moderate changes of pollutant emissions by fine-tuning on limited observations. We demonstrated this capability for August 2020 (\textit{Methods}), a period when anthropogenic activity and emissions in China were still strongly affected by the COVID-19 pandemic. Satellite analyses show that tropospheric column concentrations of  NO\textsubscript{2}, a key anthropogenic precursor of ozone, fluctuated considerably in China throughout 2020, declining during strict lock-downs and rebounding when controls eased \cite{liu2021nitrogen}. In July and August 2020, NO\textsubscript{2} column concentrations over China were approximately 20-30\% below the levels during July-August of the previous year \cite{liu2021nitrogen}. Fig.  \ref{fig:flex_adaptation} shows that AirFusion systems tuned on observations during June 2024 or during June 2020 both produced biased surface ozone forecasts across China in August 2020. In comparison, when fine-tuned with July 2020 observations, the adapted AirFusion system produced improved forecasts, with lower RMSEs (29.7 $\pm$ 6.9 $\mu$g m$^{-3}$), smaller and less scattered NMBs (3\% $\pm$ 14\%), and better temporal correlation ($r$ = 0.67 $\pm$ 0.20). The computational demand of such adaptation was minimal, requiring only 30 minutes on a single NVIDIA RTX 4090 GPU. This low computational overhead allows AirFusion to be fine-tuned frequently, potentially in near-real-time, adjusting to the rapid emission changes in China and other fast-developing regions, or during periods of abrupt emission variations.


\section{Probabilistic air quality forecasting with AirFusion}
AirFusion's computational efficiency enables probabilistic air quality forecasts driven by an ensemble of future weather scenarios, thereby explicitly quantifying the impacts of meteorological uncertainty on air quality predictions \cite{zhang2023deep}. Fig.  S4a-b show the standard deviations of MDA8O\textsubscript{3} concentrations ($\sigma_w$) across AirFusion's 30 ensemble ozone forecast members, which reflect the susceptibility of ozone forecasts to meteorological forecast errors. The spatial pattern of $\sigma_w$ was consistent with the variability of MDA8O\textsubscript{3} concentrations, with highest $\sigma_w$ values over the BTH area (>18 $\mu$g m$^{-3}$ on Day 1). These uncertainty were similar to AirFusion's ozone forecast RMSEs and exceeded China's operational forecast accuracy targets, indicating weather-induced uncertainty as a major, often dominant, source of error that must be accounted for in reliable air quality forecasting.

\begin{figure}
    \centering
    \includegraphics[width=\linewidth]{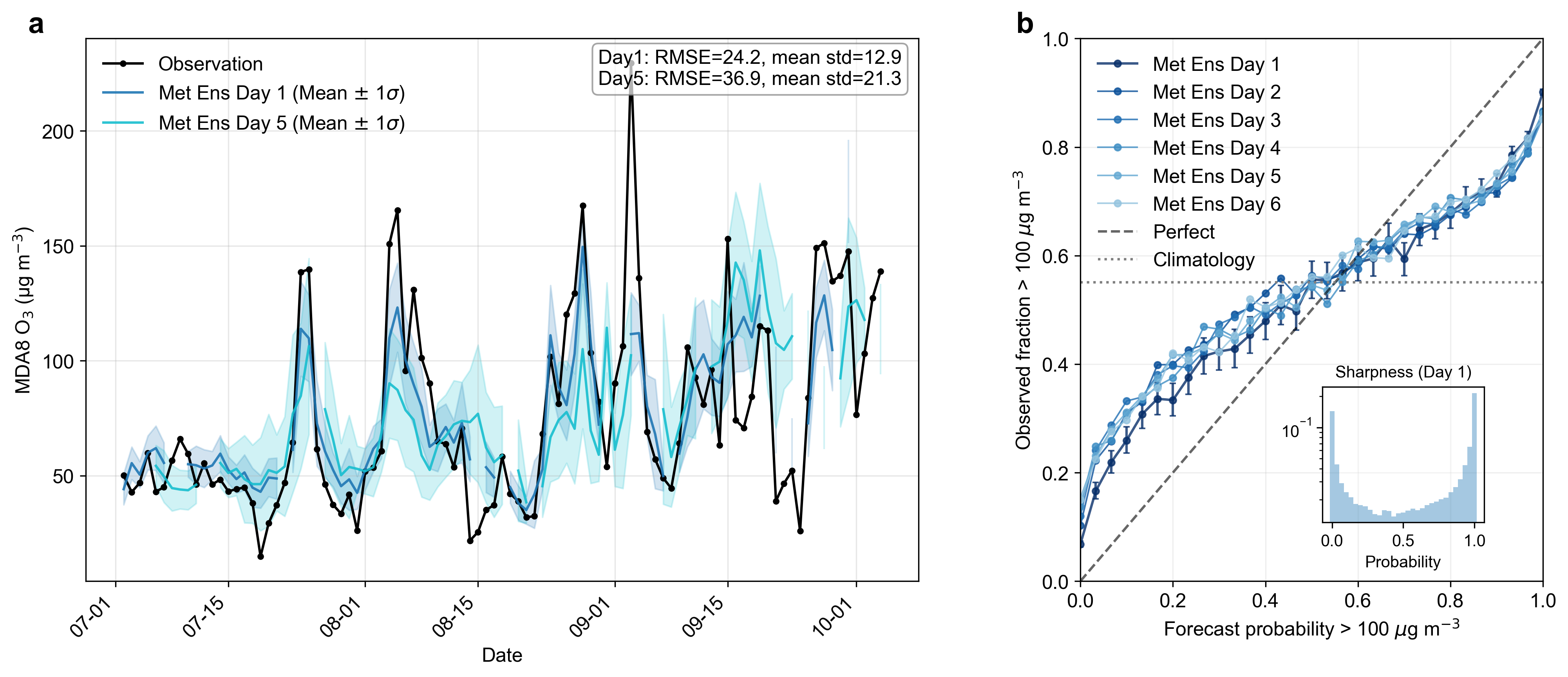}
    \caption{Probabilistic ozone forecasting with AirFusion. \textbf{a}, Time series of observed (black) and ensemble-mean forecast MDA8O\textsubscript{3} concentrations for Shenzhen during July–September 2024. The dark- and light-blue lines show Day-1 and Day-5 forecasts, respectively; shading indicates the standard deviation across ensemble members. RMSEs and standard deviations of the ensemble mean against observations are shown inset. \textbf{b}, Comparison of the ozone exceedance probability (OEP) forecast by AirFusion with the observed exceedance frequency across 341 Chinese cities during July–September 2024, using the WHO guideline (100 $\mu$g m$^{-3}$ for MDA8O\textsubscript{3}) as the threshold.} 
    \label{fig:timeseries_sz}
\end{figure}

AirFusion's ensemble-based probabilistic forecast capability is illustrated for Shenzhen (Fig. \ref{fig:timeseries_sz}a), a coastal city in southern China where air quality is highly sensitive to mesoscale wind convergence \cite{10.5194/acp-20-6305-2020}. On Day 1, the range of probable weather outcomes produced a considerable ensemble spread ($\sigma_w$ = 12.9 $\mu$g m$^{-3}$), with the ensemble mean reasonably matching the observations (RMSE of the ensemble mean MDA8O\textsubscript{3} against observations = 24.2 $\mu$g m$^{-3}$, $r$ = 0.84). By Day 5, the ensemble ozone forecast members scattered more widely ($\sigma_w$ = 21.3 $\mu$g m$^{-3}$) and the ensemble mean became more biased (RMSE = 36.9 $\mu$g m$^{-3}$, $r$ = 0.51). This growth of ozone forecast error and scatter with lead time is driven by the innate meteorological forecast uncertainty and cannot be eliminated by refining CTM representations or emission inventories.

Given the substantial and unavoidable impacts of meteorological uncertainty, we argue that forecasting the ozone exceedance probability (OEP), analogous to the precipitation probability in weather forecasts, provides more actionable information for public health protection and science-based decision-making than a single deterministic forecast. AirFusion calculates OEP as the fraction of ensemble members predicting surface ozone concentrations above a defined threshold. Fig. \ref{fig:timeseries_sz}b compares the forecast OEP against the observed frequency of ozone exceedance across 341 Chinese cities, using the World Health Organization's guideline of 100 $\mu$g m$^{-3}$ for MDA8O\textsubscript{3} concentration as threshold. AirFusion reliably predicted OEP, with a mean expected calibration error (ECE, \textit{Supplementary Information}) of 0.106 for Day 1. As discussed above, this ECE cannot be eliminated due to inherent weather forecast uncertainty. For the same reason, an unbiased single-value ozone forecast, such as that generated by a traditional CTM or a single AI-AQ inference, would, at best, correctly predict ozone exceedance at a probability approximately equal to the OEP value from AirFusion.

The performance of AirFusion's OEP forecasts also reflected the meteorological drivers of ozone pollution and their predictability. Forecast skills tended to degrade more sharply with lead time at low OEP values, whereas skills remained relatively stable with lead time at higher OEP (Fig. \ref{fig:timeseries_sz}b). This pattern is consistent with the known meteorological drivers of O\textsubscript{3} pollution in Shenzhen and other parts of Eastern China: high-OEP events are often associated with persistent, large-scale features (e.g., subsidence from the West Pacific Subtropical High or approaching typhoons), whereas low-OEP events are linked to more variable, less predictable conditions such as local wind convergence.

Due to their stoichiastic learning process, diffusion models are inherently probabilistic \cite{leutbecher2020probabilistic}. Consequently, every inference of AirFusion generates a slightly different air quality forecast, with a scatter ($\sigma_d$) driven by multiple systematic uncertainties, including errors in CTM physics and chemistry, emissions, fitting errors, observational noise, and, to some extent, the weather forecast uncertainty captured during fine-tuning. We found that $\sigma_d$ and $\sigma_w$ were highly correlated in space (spatial correlation $r$ = 0.97 and 0.93 for Day 1 and Day 3, respectively), but the $\sigma_d$ values were smaller than $\sigma_w$ (Fig. S4c-d). Therefore, the impacts of weather uncertainty can only be fully accounted for by driving AirFusion with ensemble weather forecasts.

\section{Discussion}
We developed AirFusion, a hybrid diffusion-based framework for accurate, probabilistic air quality forecasting with high efficiency. The framework uniquely integrates mechanistic insights from CTM simulations with data-driven correction fine-tuned on a small set of observations. This design enables AirFusion to deliver reliable 6-day ensemble surface ozone forecasts over China and adapt rapidly to changing emissions. A key finding of this work is that inherent uncertainty of weather forecasts imposes a substantial and unavoidable limit on air quality predictability, which is a source of error rarely addressed in traditional CTM-based forecasting. AirFusion explicitly quantifies this uncertainty through ensemble-based probabilitic diagnostics. Recognizing and quantifying this meteorological uncertainty is essential for robust, risk-informed air quality management and policy making. 

AirFusion demonstrates the viability of hybrid physics–AI systems for operational air quality forecasting and is readily extensible to other pollutants and geographical regions. Its grid-scale architecture supports future integration with advanced data assimilation. Overall, AirFusion represents a significant advance over traditional CTMs and existing AI-AQ models in forecast accuracy, spatial resolution, computational efficiency, and operational robustness, providing a powerful and adaptable tool for managing air-pollution impacts on public health and ecosystems under evolving emissions and meteorological conditions.

\section{Methods}
\subsection{Regional ozone simulation using the WRF-GC model}
We used the WRF-GC v2.0 regional air quality model \cite{lin2020wrf,feng2021wrf} to simulate surface ozone concentrations over China from 25 June to 30 September 2021; the results were used to train the AirFusion-S and AirFusion-T modules. WRF-GC v2.0 is an online coupling of the Weather Research and Forecasting (WRF) model (v3.9.1.1) \cite{jensen2021description} and the GEOS-Chem chemical transport model (v12.8.2) \cite{bey2001global}. Two sets of 27-km-resolution simulations were conducted. The first set, "WG-FNL-2021", was driven and nudged every 6 hours with the National Centers for Environmental Prediction Final Operational Global Analysis (NCEP GDAS/FNL, 0.25\textdegree~spatial resolution, \url{https://rda.ucar.edu/datasets/ds083.3/}, accessed 6 May 2022; \cite{ncep2000operational}) to closely replicate observed meteorology; the simulated ozone and meteorological fields from this set have been previously evaluated against observations \cite{zhang2023deep}. The second set, "WG-hindcast-2021", was designed to manifest surface ozone differences associated with small variations in meteorology. For each day during the study period, a 7-day hindcast was initialized from the NCEP GDAS/FNL reanalysis and run freely; results from the first simulation day were discarded. This protocol produced six model outputs for every day during the study period, each perturbed with slightly different meteorological initial conditions, thus capturing a range of realistic synoptic-scale weather patterns and mesoscale variations that influence surface ozone concentrations \cite{zhang2023deep}. 

We also used WRF-GC to forecast surface ozone concentrations across China for July to September 2024 to compare with the performance of the AirFusion system. Further details of the emissions and physical configurations in our WRF-GC simulations are given in the Supporting Information.

\subsubsection{Surface ozone measurements, down-scaled meteorological reanalysis, and ensemble weather forecasts over China}
Hourly surface ozone concentration measurements from 2027 sites across 341 Chinese cities were used to fine-tune the AirFusion-T-FT module, to generate continuous 2-D ozone concentration fields as initial conditions for AirFusion forecasts, and to evaluate forecast performance of models. These measurements were managed by the China National Environmental Monitoring Centre (\url{https://www.cnemc.cn}, last accessed: 20 Jan 2026); we applied a quality control protocol to remove errors and outliers \cite{wang2021sensitivities}.

To drive the AirFusion forecasts, we used the 30-member ensemble weather forecasts from the National Oceanic and Atmospheric Administration Global Ensemble Forecast System (NOAA GEFS, \url{https://registry.opendata.aws/noaa-gefs/}, last accessed: 26 July 2025 \cite{NOAA_GEFS}), which has a 0.25\textdegree~resolution spatial resolution and 3-hour temporal resolution \cite{NOAA_GEFS}. 

We used the WRF model to downscale National Centers for Environmental Prediction Final Operational Global Analysis (NCEP GDAS/FNL, 0.25\textdegree~spatial resolution, \url{https://rda.ucar.edu/datasets/ds083.3/}, accessed 6 May 2022; \cite{ncep2000operational}) to drive AirFusion hindcasts (AirFusion-nudgedH) over China during June-September 2020 and 2024. The WRF simulations were physically configured identically to the WRF-GC simulations, except the wind speed, temperature, humidity, and pressure were nudged toward the NCEP reanalysis every 6 hours to ensure best alignment with the reanalysis. 

\subsection{Construction and training of the AirFusion system}
The AirFusion system consists of three modules of different functionalities: AirFusion-S, AirFusion-T, and AirFusion-T-FT (Fig.  1). These modules share a common diffusion model structure, each generating a 2-D ozone concentration field by progressively denoising a noisy image. Fig.  S5 shows the common diffusion model structure, based on a modified UNet backbone \cite{karras2024analyzing} consisting of encoder blocks that down-sample the input to extract features, decoder blocks that up-sample and reconstruct the image, and skip connections that link corresponding layers to preserve spatial information. In addition, a noise embedding block applies a noise level ($\gamma$) to each encoder and decoder, and a time embedding block provides temporal information. 

We drew from our operational air quality forecast experience and selected the following hourly meteorological variables as inputs to AirFusion: surface air temperature (T2m), surface relative humidity (RH2m), winds at 10 m (U10m, V10m), 850 hPa (U850, V850), and 925 hPa (U925, V925), surface pressure (PS), 6-hour accumulated precipitation (PREC), and topography masks (mask850, mask925) to indicate the availability of meteorological data at 850 hPa and 925 hPa pressure levels. These meteorological inputs from WRF-GC (for training) and from GEFS (for fine-tuning and forecast) were regridded to the 27-km AirFusion grid using nearest-neighbor interpolation. We constructed 2-D surface ozone concentrations fields of consistent format from WRF-GC (for training) and from observations (for fine-tuning and forecast) as inputs to AirFusion. Model results from WRF-GC were sampled at locations of surface sites, then interpolated with inverse-distance weighting (IDW) onto the 27-km AirFusion grid; this process produced 2-D "pseudo" observations across China to mimic the spatial information provided by surface observations. Surface observations were similarly mapped, averaging multiple observations within the same AirFusion grid.

We trained the three AirFusion modules in two stages; all training were optimized using the the Mean Squared Error (MSE) loss function (\textit{Supplementary Information}). AirFusion-S was trained with WRF-GC 2-D pseudo observations of ozone and meteorological fields at the previous ($t-\Delta t$) and present ($t$) time steps as input, and the WRF-GC continuous 2-D pollution concentration image at time $t$ as output. 
AirFusion-T was trained with WRF-GC 2-D concentrations ($t = t-\Delta t$ and $t$) and meteorology (at $t = t$ and $t+\Delta t$) as input, and the WRF-GC 2-D concentrations at time $t+\Delta t$ as output. In this way, AirFusion-S learned from WRF-GC how spatially sparse concentrations are connected across the domain, while AirFusion-T learned from WRF-GC how pollution-meteorology relationship advance pollution concentrations in time. We chronologically split the WRF-GC simulation data into training (first 70\%) and testing (later 30\%) datasets. Subsequently, the pre-trained AirFusion-T was fine-tuned with hourly surface ozone observations GEFS ensemble mean weather forecasts during June to September 2021, producing AirFusion-T-FT in which the neural network weights were adjusted to remove biases from WRF-GC and the weather forecast model. 

\section*{Data availability}
Hourly observations of surface ozone concentrations across China were downloaded from \url{https://www.cnemc.cn}. NCEP GFES outputs were downloaded daily from \url{https://registry.opendata.aws/noaa-gefs/}. NCEP FNL datasets were downloaded from \url{https://rda.ucar.edu/datasets/ds083.3/}.

\section*{Code availability}
The code for the operational AirFusion model will be made publicly available upon publication. WRF-GC code is available at \url{https://github.com/WRF-GC/wrf-gc-release}.

\section*{Author contribution}
A.D., A.Z., and T.M.F. conceived of the study, conducted the model development and analyses, and wrote the paper. A.Z., T.M.F., Y.H., and Q.C. supervised the project. Y.C., J.M., W.T., W.C.C., L.Z., X.Y., and G.B. provided constructive suggestions to improve this study. All authors gave feedback and contributed to editing the paper.

\section*{Competing interests}
The authors declare no competing interests. 


\section*{Acknowledgements}
This work was supported by the National Natural Science Foundation of China (42325504, 42461160326, 42305188), the Guangdong Provincial Field Observation and Research Station for Coastal Atmosphere and Climate of the Greater Bay Area (2021B1212050024), the Shenzhen Science and Technology Program (KQTD20210811090048025), the High-level Special Funds (G03034K006), and Hong Kong Research Grants Council (Grant Nos. 15223221, 15219722, and 25231725). Computational resources were supported by the Center for Computational Science and Engineering of the Southern University of Science and Technology.

\bibliography{sn-bibliography}

\end{document}


\title[Article Title]{\centering \textit{Supplementary Information} \\ \textit{for} \\ Diffusion-based Probabilistic Air Quality Forecasting with Mechanistic Insight}


\author[1,2,3,4]{\fnm{Ao} \sur{Ding}}\email{ao2024.ding@connect.polyu.hk}
\author*[3,4]{\fnm{Aoxing} \sur{Zhang}}\email{zhangax@sustech.edu.cn}
\author*[3,4,5]{\fnm{Tzung-May} \sur{Fu}}\email{fuzm@sustech.edu.cn}
\author*[1]{\fnm{Yuanlong} \sur{Huang}}\email{yhuang@eitech.edu.cn}
\author*[2]{\fnm{Qianjie} \sur{Chen}}\email{qianjie.chen@polyu.edu.hk}
\author[3,4]{\fnm{Yuyang} \sur{Chen}}
\author[3,4]{\fnm{Jiajia} \sur{Mo}}
\author[3,4]{\fnm{Wei} \sur{Tao}}
\author[3,4]{\fnm{Wai-Chi} \sur{Cheng}}
\author[3,4]{\fnm{Lei} \sur{Zhu}}
\author[3,4]{\fnm{Xin} \sur{Yang}}
\author[6]{\fnm{Guy} \sur{Brasseur}}

\affil[1]{School of the Environment and Sustainable Engineering, Eastern Institute of Technology, Ningbo, Zhejiang, 315200, China}
\affil[2]{Department of Civil and Environmental Engineering, The Hong Kong Polytechnic University, Hong Kong SAR, China}
\affil[3]{State Key Laboratory of Soil Pollution Control and Safety, Shenzhen Key Laboratory of Precision Measurement and Early Warning Technology for Urban Environmental Health Risks, School of Environmental Science and Engineering, Southern University of Science and Technology, Shenzhen, Guangdong, 518055, China}
\affil[4]{Guangdong Provincial Field Observation and Research Station for Coastal Atmosphere and Climate of the Greater Bay Area, Southern University of Science and Technology, Shenzhen, Guangdong, 518055, China}
\affil[5]{National Center for Applied Mathematics Shenzhen, Shenzhen, Guangdong, 518055, China}
\affil[6]{Max Planck Institute for Meteorology, Hamburg, 20146, Germany}

\maketitle

\section{Supplementary Methods}
\subsection{WRF-GC simulations and WRF meteorological down-scaling}
\label{method:WRF-GC}
Our WRF-GC simulations were driven by monthly anthropogenic emissions from the Multi-resolution Emission Inventory for China (MEIC), developed for the year 2019 at a native spatial resolution of 0.25$^\circ$ \cite{li2019anthropogenic, li2017anthropogenic, zheng2018trends, zheng2021changes}. MEIC included emissions from industry, power generation, transportation, residential activities, and agriculture. Emissions of biogenic volatile organic compounds and soil- and lightning-NO\textsubscript{x} were calculated online in WRF-GC \citep{guenther2012model,hudman2010interannual,murray2012optimized}. 

Table \ref{tab:physical-schemes} summarizes the physical schemes used in our WRF-GC simulations. Daily chemical initial and boundary conditions to the WRF-GC model domain were from a global simulation using the Model for Ozone and Related chemical Tracers (MOZART) \citep{emmons2010description} for the year 2017, with a spatial resolution of $1.9^\circ \times 2.5^\circ$. The time steps for physical and chemical calculations were 2 minutes and 10 minutes, respectively. In the nudged hindcasts ("WG-FNL-2021" simulations), temperature, pressure, wind, and specific humidity above the boundary layer in WRF-GC were nudged towards the NCEP GDAS/FNL reanalysis every 6 hours using a four-dimensional data assimilation (nudging coefficient 0.003) \citep{stauffer1990use}.  

To train and drive AirFusion hindcasts, we also used the WRF model \cite{jensen2021description} to downscale the NCEP GDAS/FNL renalysis to AirFusion's temporal (3-hourly) and spatial (27 km) resolution. The nudging settings were identical to the nudging of WRF-GC described above.

\subsection{Diffusion model architecture}
\label{method:diffusion}
Our AirFusion modules adopt the diffusion framework proposed by Karras et al. \cite{karras2022elucidating} as implemented in the open-source project denoising-diffusion-pytorch (\url{https://github.com/lucidrains/denoising-diffusion-pytorch}, last accessed: 1 June 2025, forked to \url{https://github.com/topleo/denoising-diffusion-pytorch}) with a modified UNet as the backbone to processes heterogeneous data. 

\begin{figure}
    \centering
    \includegraphics[width=\linewidth]{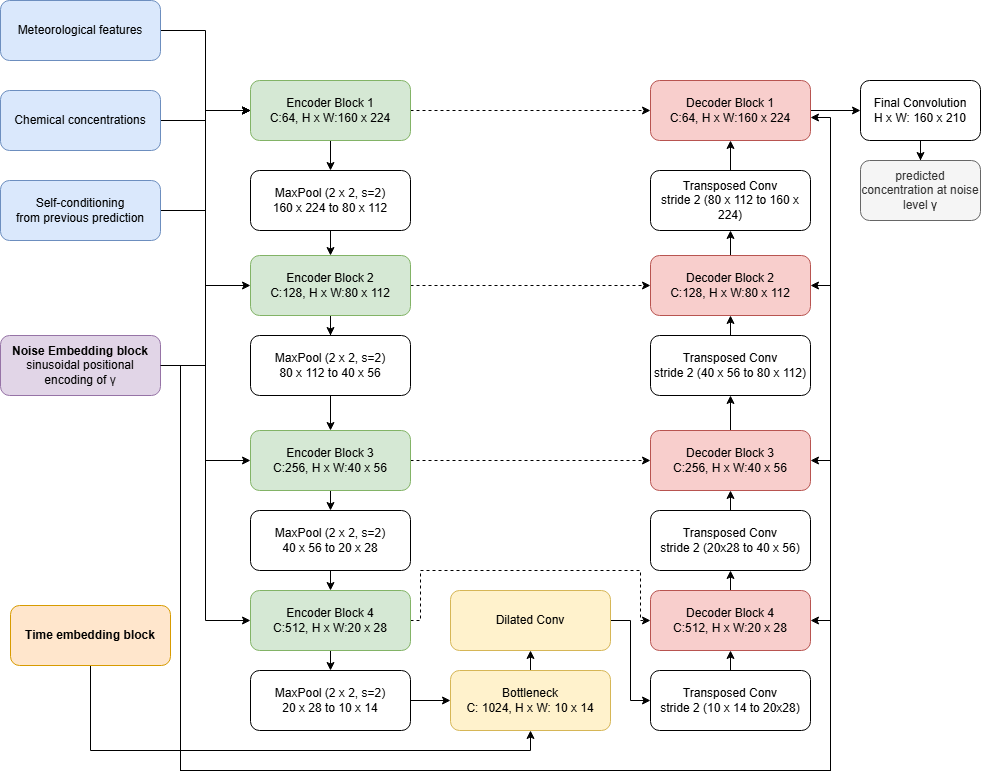}
    \caption{Network structure of the denoiser in the AirFusion modules.}
    \label{fig:denoiser}
\end{figure}

Fig. \ref{fig:denoiser} illustrates the denoiser network. Blue boxes represent 2-D data matrices that serve as direct input to the model. The purple box corresponds to the noise embedding block, where the noise level $\gamma$  is transformed as $c_{\text{noise}}=\log(\gamma)\times0.25$. The orange box is the time embedding block, where the hour-of-day $t$ is transformed to $t_{\sin}=\sin(2*\pi*t/24)$ and $t_{\cos}=\cos(2*\pi*t/24)$, and the day-of-week $d$ was transformed to $d_{\sin}=\sin(2*\pi*d/7)$ and $d_{\cos}=\cos(2*\pi*d/7)$.

The encoding path proceeds through with four successive encoding blocks: Encoder Block 1 (C=64, H$\times$ W=160$\times$224), Encoder Block 2 (C=128, H$\times$W=80$\times$112), Encoder Block 3 (C=256, H$\times$W=40$\times$56), and Encoder Block 4 (C=512, H$\times$W=20$\times$28. Each encoder block is followed by max-pooling for down-sampling, culminating in a final spatial dimention of  10×14. Conditioning signals are injected via specialized embedding blocks along the path: the noise embedding block provides sinusoidal positional encodings based on the denoising stage, while the time embedding block introduces temporal context to capture sequential dependencies in the atmospheric processes.

At the core of the network, a bottleneck layer employs a dilated convolution (C=1024, H$\times$W=10$\times$14) to expand the receptive field, thereby capturing the broader mesoscale interactions (such as pollutant transport) without altering spatial dimensions. The decoder then mirrors the encoder's structure. Beginning with Decoder Block 4 (C=512, H$\times$W=20$\times$28), which is reached via a transposed convolution with stride 2 that up-samples the bottleneck output from 10$\times$14 output to 20$\times$28. Each subsequent decoder blocks are similarly preceded with a transponsed convolution block (stride 2) for up-sampling. Dashed lines in Fig.  S5 indicate skip connections that link corresponding encoder and decoder blocks at matching resolutions, facilitating the retention of high-resolution details otherwise lost during down-sampling. Finally, the denoiser output 2-D maps of pollutant concentration at noise level $\gamma$. The noise level $\gamma$ is randomly sampled during training, and $\gamma$ is sampled using DPM-Solver++\citep{lu2022dpmpp} solver for the denoising processes.

\subsection{Loss function}
\label{method:loss_function}
All AirFusion training were optimized using the the Mean Squared Error (MSE) loss function: 
\begin{equation}
\mathcal{L}_{\text{MSE}}(\hat{y}, y) = \frac{1}{n} \sum_{i=1}^{n} (\hat{y}_i - y_i)^2
\end{equation}
    $\hat{y}_i$ is the predicted value of the $i$\textsuperscript{th} sample, $y_i$ is the ground truth of the $i$\textsuperscript{th} sample. For training AirFusion-S and AirFusion-T, the ground truth was the 2-D WRF-GC simulated ozone concentration field; for training AirFusion-T-FT, the ground truth was the 2-D interpolated observed ozone concentration field.

\subsection{Training strategy }
AirFusion-S and AirFusion-T were trained using WRF-GC simulated meteorological fields and simulated pollutant concentrations sampled at locations of Chinese surface measurement sites, managed by the China National Environmental Monitoring Centre (CNEMC, \url{http://www.cnemc.cn}, last accessed: 5 May 2022). AirFusion-T-FT was trained using ensemble mean weather forecasts from the National Oceanic and Atmospheric Administration Global Ensemble Forecast System (NOAA GEFS, \url{https://registry.opendata.aws/noaa-gefs/}, last accessed: 26 July 2025 \citep{NOAA_GEFS}) and the surface pollutant observations at the CNEMC sites. Before inputting to the denoiser, the spatially discrete surface observations and "psuedo-observations" from WRF-GC were interpolated using Inverse Distance Weighting (IDW) to generate continuous 2D fields. A time-dependent site mask was applied to indicate availability of sites and valid measurements. Temporal information, including the hour-of-day and the day-of-week, was provided to capture temporal variations. 

For training AirFusion-S and AirFusion-T, the learning rate was set to $10^{-4}$ and decaying by half every 200 epochs for a total of 600 epochs. The Adam optimizer was employed to improve model convergence and stability\cite{kingma2014adam}. For fine-tuning AirFusion-T-FT, the learning rate was set to $10^{-5}$ for a total epoch of 10 to avoid overfitting and to retain mechanistic insights from WRF-GC.

\subsection{Wavelet scale-decomposition of differences in O\textsubscript{3} forecasts}
To understand how AirFusion learned the latent meteorology-pollution relationship from WRF-GC and from fine-tuning, we used 2-D discrete wavelet transformation (DWT) to decompose the spatial scales of the differences in forecast O\textsubscript{3} concentrations from WRF-GC, AirFusion-noFT, and the fully fine-tuned AirFusion. The DWT was conducted using the python function \verb|pywt.wavedec2| (Daubechies-4 wavelet, decomposition level $L=4$).

\subsection{Reliability diagram and expected calibration error (ECE)}
\label{method:ECE}
Reliability diagram is a standard tool to visualize the performance of a probabilistic classifier. The expected calibration error (ECE) is a scaler diagnostic that evaluates how well a machine learning model's predicted probabilities match the actual observed frequencies over the full range of probabilities\citep{naeini2015obtaining}. 

We diagnosed the ECE of AirFusion's ozone exceedance probability (OEP) forecast skills by comparing the forecast OEP to the actual observed O\textsubscript{3} exceedance/non-exceedance across 341 Chinese cities for July to September 2024 (92 days). $N$ is the total number of OEP forecast samples; in our case $N =$ 341 (cities) $\times$ 92 (days) = 31,372. For the $i$\textsuperscript{th} sample ($i=$ 1 to $N$), the maximum estimated probability, $\hat{p}_i$, is the probability of exceedance (OEP$_i$) or non-exceedance ($1-$ OEP$_i$), whichever is >0.5.
\begin{equation}
    \hat{p}_i = \max(\text{OEP}_i, 1-\text{OEP}_i)
\end{equation}

We divide the $N$ samples of OEP forecasts according to their $\hat{p}_i$ into $M=30$ bins between 0 and 1. The ECE is calculated as:

\begin{equation}
\text{ECE} = \sum_{m=1}^M \frac{|\mathbf{B}_m|}{N} \left| \text{acc}(\mathbf{B}_m) - \text{conf}(\mathbf{B}_m) \right| 
\end{equation}
where $\mathbf{B}_m$ is the set of samples from all cities and forecast days falling into the $m$\textsuperscript{th} (= 1 to $M$) probability bin; $|\mathbf{B}_m|$ is the size of $\mathbf{B}_m$.

The function conf($\mathbf{B}_m$) calculates the average $\hat{p}_i$ within the $\mathbf{B}_m$ set:
\begin{equation}
\text{conf}(\mathbf{B}_m) = \frac{1}{|\mathbf{B}_m|} \sum_{i \in \mathbf{B}_m} \hat{p}_i 
\end{equation}

The function acc($\mathbf{B}_m$) calculates the fraction of times of correct categorical prediction, in the $m$\textsuperscript{th} probability bin. In our case, if the forecast OEP was >0.5 and an O\textsubscript{3} exceedance actually occurred, or if the forecast OEP was <0.5 and an O\textsubscript{3} exceedance did not occur, then $\mathbf{1}(\hat{y}_i = y_i) = 1$
\begin{equation}
\text{acc}(\mathbf{B}_m) = \frac{1}{|\mathbf{B}_m|} \sum_{i \in \mathbf{B}_m} \mathbf{1}(\hat{y}_i = y_i) 
\end{equation}
where $\hat{y}_i$ is predicted exceedance/non-exceedance, and $y_i$ is the actual exceedance/non-exceedance.
\newpage
\section{Supplementary Information}

The data used for fine-tuning and forecasting are obtained from publicly available sources, including:
\begin{itemize}
    \item \textbf{GEFS} (Global Ensemble Forecast System): 0.25\textdegree spatial resolution and 3-hour temporal resolution, initialized at 00, 06, 12, and 18 UTC daily. The full set of 30 ensemble members (used in this study) can be downloaded from \url{https://nomads.ncep.noaa.gov/gribfilter.php?ds=gefs_atmos_0p25s}. An archived 20-member ensemble version is available at \url{https://registry.opendata.aws/noaa-gefs/}. 

    \item \textbf{NCEP GDAS/FNL}: NCEP FNL (Final) operational global analysis with 0.25\textdegree spatial resolution, available at 00, 06, 12, and 18 UTC each day. The dataset can be downloaded from \url{https://gdex.ucar.edu/datasets/d083003/}.
 
    \item \textbf{Surface O\textsubscript{3} observations}: The original monitoring data are published by the China National Environmental Monitoring Centre (CNEMC) at \url{https://air.cnemc.cn:18007/}. The archived dataset used in this study is obtained from \url{https://quotsoft.net/air/}. 
\end{itemize}

\newpage

\section{Supplementary Tables}

\begin{table}[htbp]
\centering
\caption{Physical schemes used in the WRF-GC simulations}
\label{tab:physical-schemes}
\begin{tabular}{p{3.2cm} p{4.2cm} p{4.2cm}}
\toprule
Physical Process & Scheme & References \\
\midrule
Microphysics 
& Morrison 2-moment bulk microphysics 
& Morrison et al. (2009)\citep{morrison2009impact} \\

Radiative transfer 
& RRTMG 
& Iacono et al. (2008)\citep{iacono2008radiative}; 
Mlawer et al. (1997)\citep{mlawer1997radiative}; 
Heald et al. (2014)\citep{heald2014contrasting} \\

Surface-layer process 
& Monin-Obukhov similarity scheme 
& Monin \& Obukhov (1954)\citep{monin1954basic} \\

Land surface module 
& Noah land surface model 
& Chen et al. (1996)\citep{chen1996modeling} \\

Boundary-layer physics 
& MYNN 2.5 level TKE scheme 
& Mellor \& Yamada (1982)\citep{mellor1982development}; 
Nakanishi \& Niino (2009)\citep{nakanishi2009development} \\

Cumulus scheme 
& New-Tiedtke scheme 
& Tiedtke (1989)\citep{tiedtke1989comprehensive}; 
Zhang et al. (2011)\citep{zhang2011improved} \\

\bottomrule
\end{tabular}
\end{table}

\newpage
\section{Supplementary Figures}
\begin{figure}
    \centering
    \includegraphics[width=\linewidth]{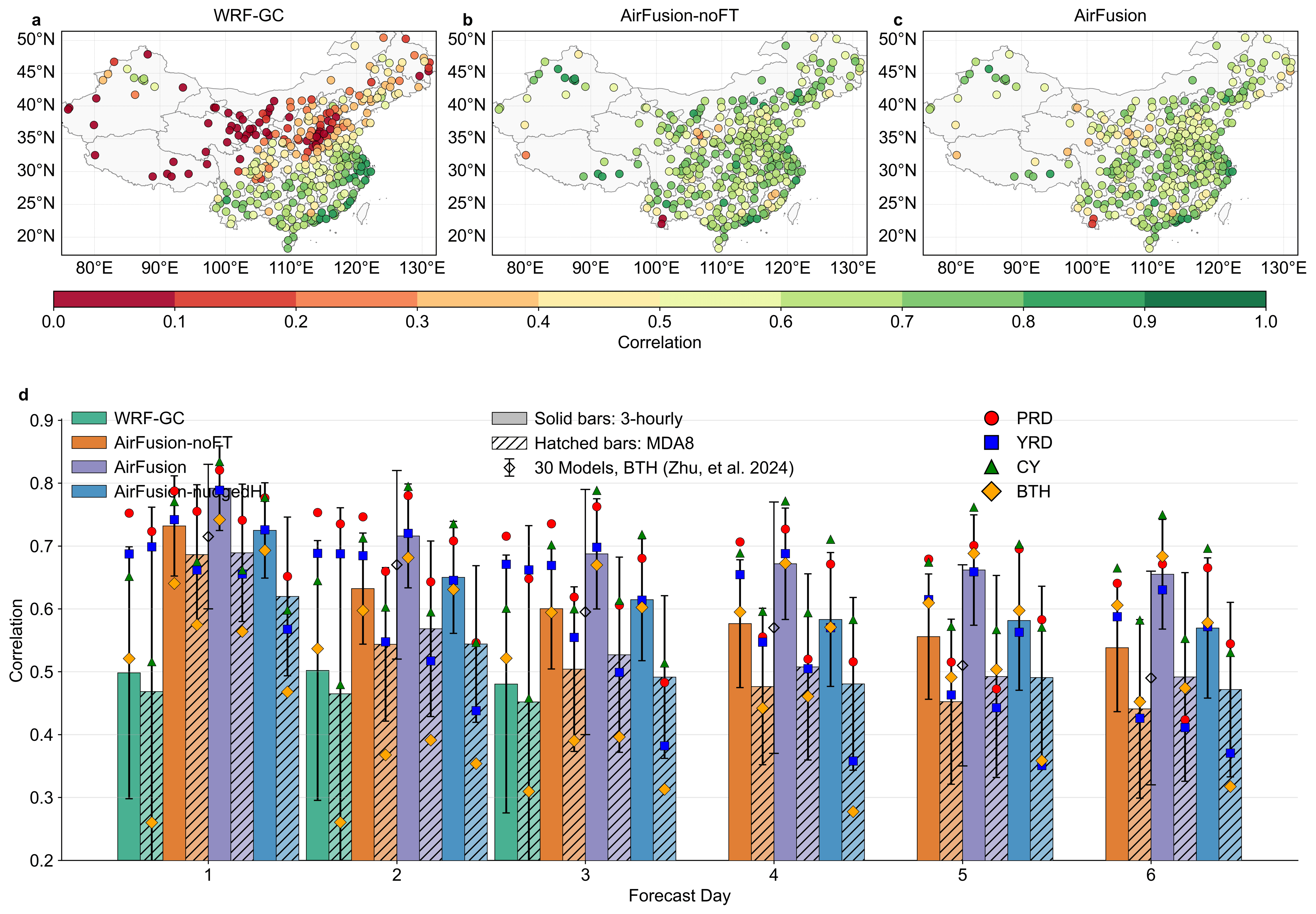}
    \caption{\textbf{Correlation of the AirFusion system in forecasting surface ozone concentrations across China.} \textbf{a-c}, Temporal correlation coefficients ($r$) of Day-1 MDA8O\textsubscript{3} forecasts for July to September 2024 from three models: WRF-GC (\textbf{a}), AirFusion without fine-tuning (AirFusion-noFT) (\textbf{b}), and the fully fine-tuned AirFusion system (\textbf{c}). MDA8O\textsubscript{3} forecasts in AirFusion-noFT and AirFusion are derived by linearly interpolating their respective 3-hourly ensemble mean ozone forecasts. \textbf{d}, Comparisons of correlation coefficients ($r$) of forecasted MDA8O\textsubscript{3} (solid bars) and 3-hourly ozone (hatched bars) concentrations averaged across Chinese cities at Day 1 to Day 6 lead times. Green, WRF-GC; orange, AirFusion-noFT; purple, AirFusion; blue, AirFusion-nudgedH. Black whiskers indicate standard deviations across Chinese cities. Also shown are model performance over four major megacity clusters (colored symbols): Beijing–Tianjin–Hebei (BTH), Yangtze River Delta (YRD), Pearl River Delta (PRD), and Sichuan–Chongqing (CY), along with the range of RMSEs from previous multi-model assessments over the BTH \citep{zhu2024comparative} (hallow rhombus)}
    \label{fig:map_r}
\end{figure}
\newpage
\begin{figure}
    \centering
    \includegraphics[width=\linewidth]{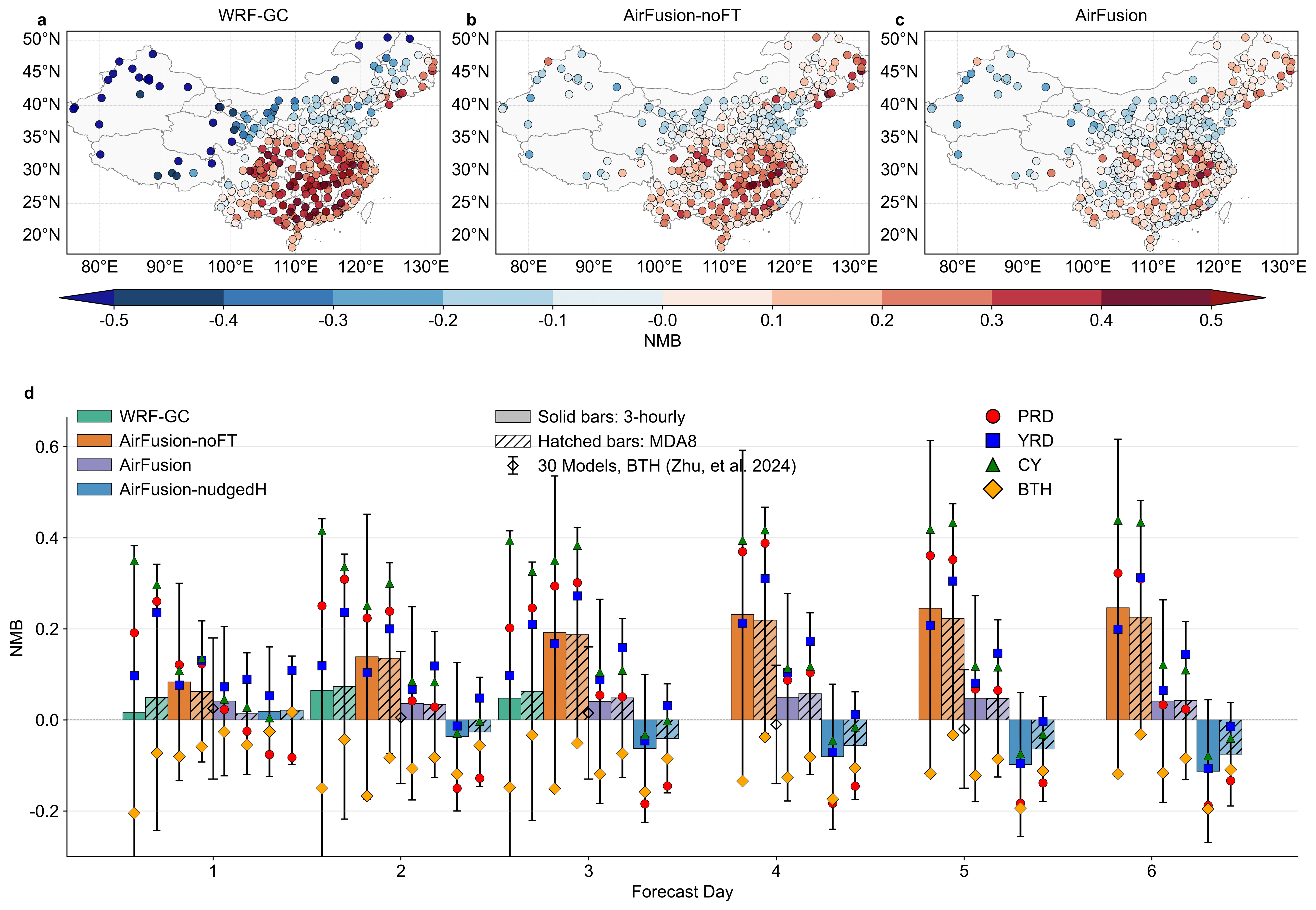}
    \caption{\textbf{Normalized Mean Bias (NMB) of the AirFusion system in forecasting surface ozone concentrations across China.} \textbf{a-c}, NMB of Day-1 MDA8O\textsubscript{3} forecasts for July to September 2024 from three models: WRF-GC (\textbf{a}), AirFusion without fine-tuning (AirFusion-noFT) (\textbf{b}), and the fully fine-tuned AirFusion system (\textbf{c}). MDA8O\textsubscript{3} forecasts in AirFusion-noFT and AirFusion are derived by linearly interpolating their respective 3-hourly ensemble mean ozone forecasts. \textbf{d}, Comparisons of NMBs of forecasted MDA8O\textsubscript{3} (solid bars) and 3-hourly ozone (hatched bars) concentrations averaged across Chinese cities at Day 1 to Day 6 lead times. Green, WRF-GC; orange, AirFusion-noFT; purple, AirFusion; blue, AirFusion-nudgedH. Black whiskers indicate standard deviations across Chinese cities. Also shown are model performance over four major megacity clusters (colored symbols): Beijing–Tianjin–Hebei (BTH), Yangtze River Delta (YRD), Pearl River Delta (PRD), and Sichuan–Chongqing (CY), along with the range of RMSEs from previous multi-model assessments over the BTH \citep{zhu2024comparative} (black diamond and whiskers)}
    \label{fig:map_nmb}
\end{figure}
\newpage
\begin{figure}
    \centering
    \includegraphics[width=1\linewidth]{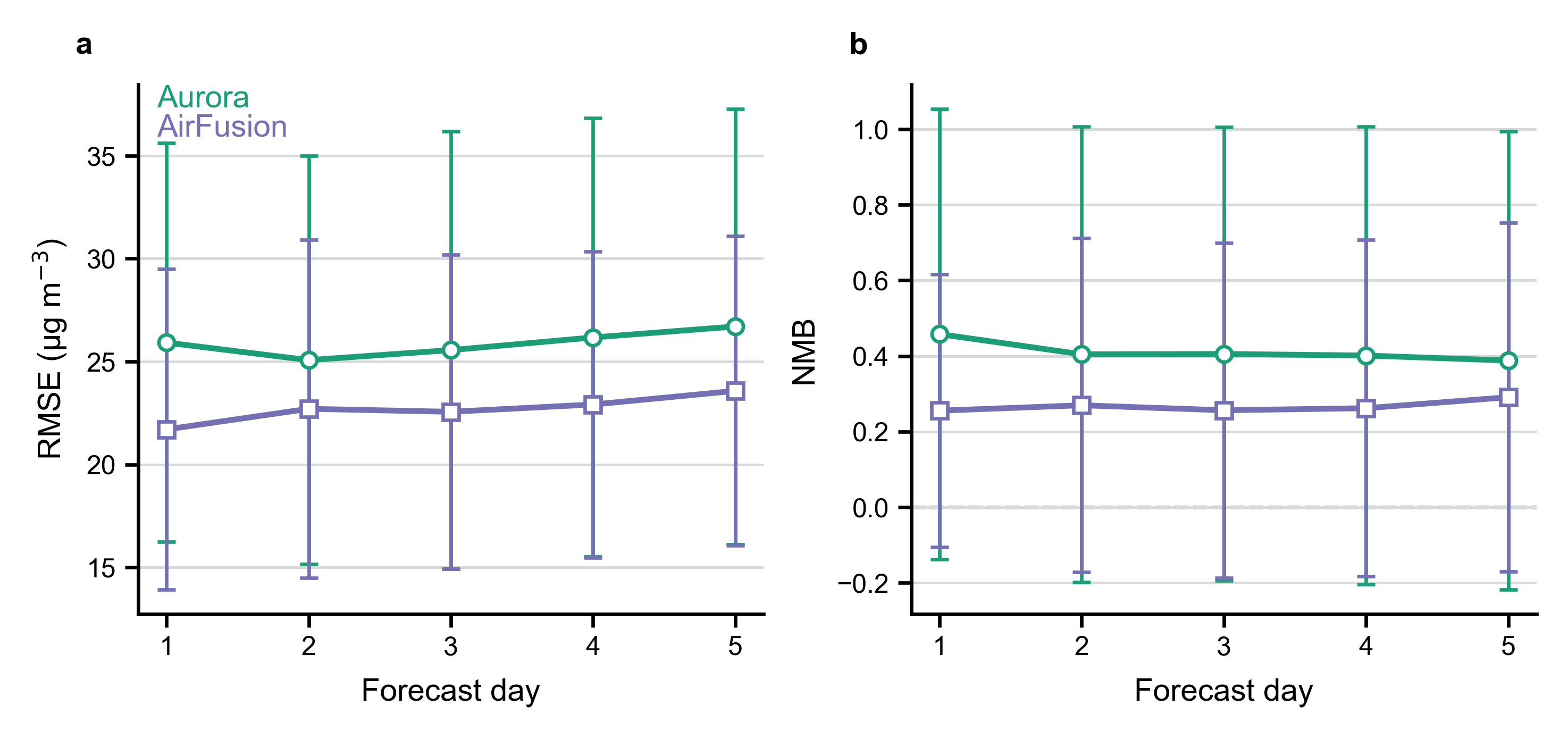}
    \caption{Performance comparison between Aurora's 00 UTC hourly O\textsubscript{3} concentration hindcast and AirFUsion's 00 UTC hourly O\textsubscript{3} concentration forecast across Chinese cities for July 2024 at different forecast lead time. \textbf{a}, RMSEs. \textbf{b} NMBs.}
    \label{fig:aurora}
\end{figure}
\begin{figure}
    \centering
    \includegraphics[width=\linewidth]{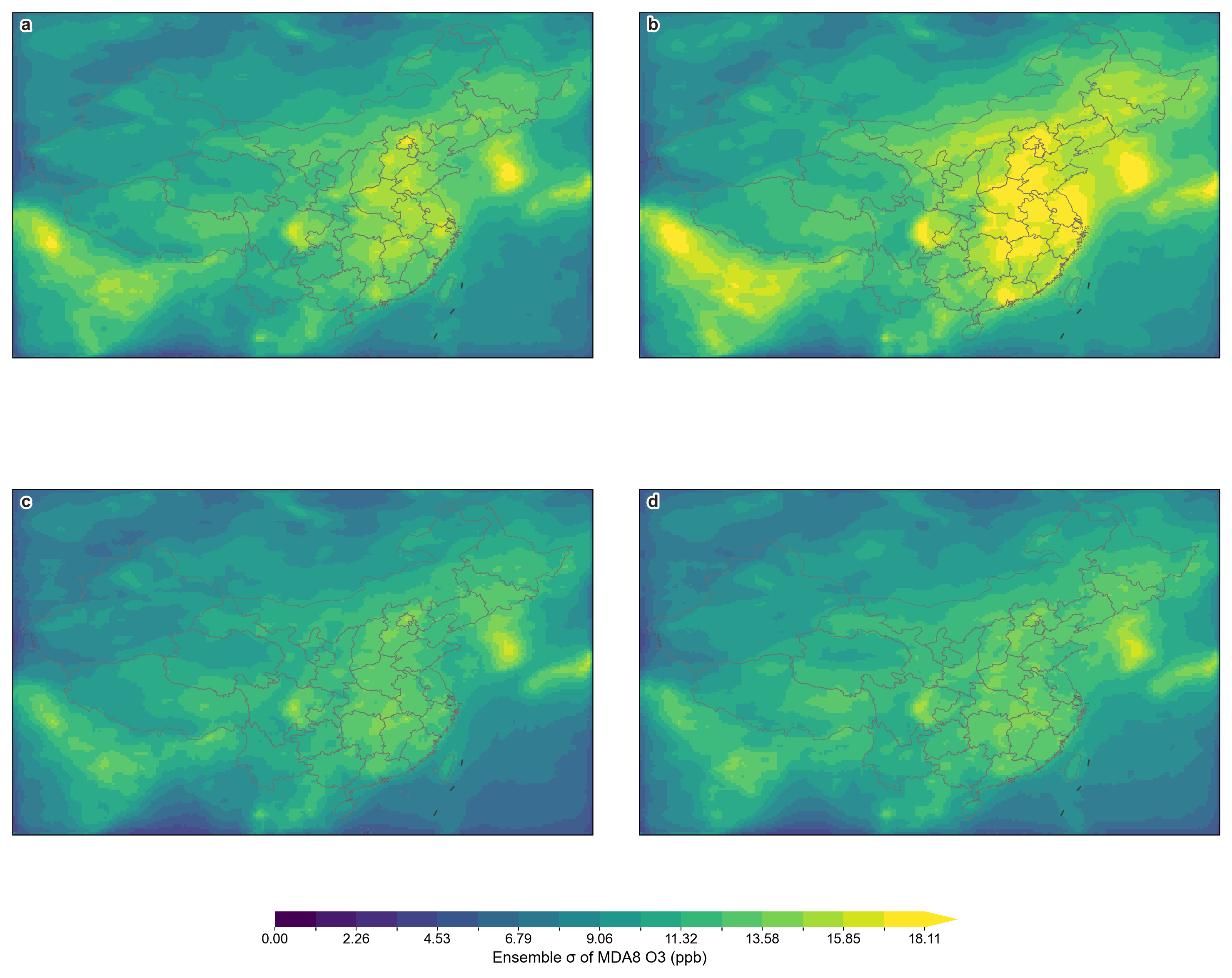}
    \caption{\textbf{a-b}, standard deviations ($\sigma_w$) of AirFusion's 30-member ensemble MDA8O\textsubscript{3} concentration forecast driven by ensemble weather forecast for Day 1 (\textbf{a}) and Day 3 (\textbf{b}). \textbf{c-d}, standard deviations ($\sigma_d$) in 30 inferences of AirFusion's MDA8O\textsubscript{3} concentration forecast for Day 1 (\textbf{c}) and Day 3 (\textbf{d}). $\sigma_d$ manifests the inherent stoichiatic property of the diffusion model.}
    \label{fig:ensemble_sigma}
\end{figure}
\newpage

\clearpage
\bibliography{sn-bibliography}